\documentclass[aps,prl,10pt,twocolumn,showpacs,preprintnumbers,amsmath,amssymb,superscriptaddress]{revtex4-1}

\usepackage[caption=false]{subfig}
\usepackage{lineno}
\usepackage{graphicx}  
\usepackage{dcolumn}   
\usepackage{bm}        
\usepackage{amssymb}   
\usepackage{amsmath}
\usepackage{blkarray, multirow, graphicx, diagbox, color, xcolor, colortbl}
\usepackage{bbm, bbold}
\usepackage{ifthen}
\usepackage{xkeyval}
\usepackage{moreverb}
\usepackage{rotating}
\usepackage{slashbox}
\usepackage{xspace}
\usepackage{nicefrac}
\usepackage[]{units}
\usepackage[]{natbib}
\usepackage{physics}
\usepackage{braket}
\usepackage{hyphenat}
\usepackage{amsbsy}

\usepackage[customcolors]{hf-tikz} 
\usepackage{tikz}
\usepackage{calc}
\usetikzlibrary{external}
\usepackage{pgffor}
\usepackage{pgfplots}
\pgfplotsset{compat=1.13}
\usepackage{pgfplotstable}
\usepgfplotslibrary{groupplots}
\usepgfplotslibrary{fillbetween}

\usetikzlibrary
{
	calc,
	decorations,
	plotmarks,
	patterns,
	positioning,
	petri,
	arrows,
	decorations.markings,
	backgrounds,
	fit,
	graphs,
	shapes.geometric,
	decorations.pathmorphing,
	shapes.misc,
	shapes,
	tikzmark,
	pgfplots.colorbrewer
}

\pgfplotsset{
        cycle from colormap manual style/.style={
            x=3cm,y=10pt,ytick=\empty,
            stack plots=y,
            every axis plot/.style={line width=2pt},
        },
}

\definecolor{colorA}{rgb} {0.58,0,0.8275}
\definecolor{colorB}{rgb} {0.11,0.663,0.51}
\definecolor{colorC}{rgb} {0.3373,0.7059,0.9137}
\definecolor{colorD}{rgb} {0.902,0.6235,0}
\definecolor{colorE}{rgb} {0.9451,0.902,0.3255}
\definecolor{colorF}{rgb} {0.3373,0.3255,0.902}
\definecolor{colorG}{rgb} {0.9451,0.3255,0.3373}

\usepackage[acronym,nohypertypes={acronym,notation}]{glossaries}
\newacronym{1D}{1D}{one\hyp dimensional}
\newacronym[shortplural={MPS}]{MPS}{MPS}{matrix\hyp product state}
\newacronym{MPO}{MPO}{matrix\hyp product operator}
\newacronym{SVD}{SVD}{singular\hyp value decomposition}
\newacronym{QCS}{QCS}{quantum\hyp computer simulator}
\newacronym{QC}{QC}{quantum computer}
\newacronym{FSM}{FSM}{finite\hyp state machine}
\newacronym{ACA}{ACA}{adaptive cross\hyp approximation}
\newacronym{CDW}{CDW}{charge\hyp density wave}
\newacronym{SDW}{SDW}{spin\hyp density wave}
\newacronym{ARPES}{ARPES}{angle-resolved photoemission spectroscopy}
\newacronym{OBC}{OBC}{open-boundary conditions}
\newacronym{PBC}{PBC}{periodic-boundary conditions}
\newacronym{TEBD}{TEBD}{time-evolution block-decimation}
\newacronym{TDVP}{TDVP}{time\hyp dependent variational principle}
\newacronym{iff}{iff}{if and only if}
\newacronym{DFT}{DFT}{density\hyp functional theory}
\newacronym{DMFT}{DMFT}{dynamical mean\hyp field theory}
\newacronym{DMRG}{DMRG}{density\hyp matrix renormalization group}
\newacronym{QMC}{QMC}{quantum Monte Carlo}
\newacronym{AIM}{AIM}{Anderson impurity model}
\newacronym{SIAM}{SIAM}{single impurity Anderson model}
\newacronym{LDA}{LDA}{local\hyp density approximation}
\newacronym{LBNL}{LBNL}{Lawrence Berkeley National Laboratory}
\newacronym{VQE}{VQE}{variational\hyp quantum eigensolver}
\newacronym{ED}{ED}{exact diagonalization}
\newacronym{QPT}{QPT}{quantum phase transition}
\newacronym{QCP}{QCP}{quantum critical point}
\newacronym{ETH}{ETH}{eigenstate thermalization hypothesis}
\newacronym{EHM}{EHM}{extended Hubbard model}
\newacronym{AKLT}{AKLT}{Affleck\hyp Lieb\hyp Kennedy\hyp Tasaki}
\newglossaryentry{QR}{name={QR},description={QR decomposition}}
\newacronym{TNS}{TNS}{tensor\hyp network state}
\newacronym{SM}{SM}{supplemental material}
\newacronym{NOO}{NOO}{natural orbital occupation}
\newacronym{NO}{NO}{natural orbital}
\newacronym{LRO}{LRO}{long\hyp range order}
\newacronym{qLRO}{qLRO}{quasi\hyp long\hyp range order}
\newacronym{SC}{SC}{Superconductivity}
\newacronym{tr-ARPES}{tr-ARPES}{time- and angle-resolved photoemission spectroscopy}

\newboolean{buildtikzpics}
\setboolean{buildtikzpics}{true}
\newif\ifrebuildtikz
\newif\ifChangeMode
\ChangeModetrue
\ChangeModefalse
\ifthenelse{\boolean{buildtikzpics}}
{
	\rebuildtikztrue
	\usetikzlibrary{external}
	\tikzexternalize[optimize=false,prefix=figures/autogen/]
}
{
	\rebuildtikzfalse
}

\usepackage{ulem}
\usepackage{cleveref}
\Crefname{appendix}{Appendix}{Appendices}
\Crefname{equation}{Equation}{Equations}
\Crefname{figure}{Figure}{Figures}
\Crefname{section}{Section}{Sections}
\Crefname{tabular}{Tabular}{Tabulars}
\crefname{appendix}{App.}{Apps.}
\crefname{equation}{Eq.}{Eqs.}
\crefname{figure}{Fig.}{Figs.}
\crefname{section}{Sec.}{Secs.}
\crefname{tabular}{Tab.}{Tabs.}

\begin{document}

\title{Laser pulse driven control of charge and spin order in the two-dimensional Kondo lattice}

\author{Benedikt Fauseweh}
\email{fauseweh@lanl.gov}
\affiliation{Theoretical Division, Los Alamos National Laboratory, Los Alamos, New Mexico 87545, USA}

\author{Jian-Xin Zhu}
\email{jxzhu@lanl.gov}
\affiliation{Theoretical Division, Los Alamos National Laboratory, Los Alamos, New Mexico 87545, USA}
\affiliation{Center for Integrated Nanotechnologies, Los Alamos National Laboratory, Los Alamos, New Mexico 87545, USA}

\date{\today}

\begin{abstract}
Fast and dynamical control of quantum phases is highly desired in the application of quantum phenomena in devices. 
On the experimental side, ultrafast laser pulses provide an ideal platform to induce femtosecond dynamics in a variety of materials.  Here, we show that a laser pulse driven heavy fermion system, can be tuned to dynamically evolve into a phase, which is not present in equilibrium. Using the state of the art time-dependent variational Monte Carlo method, we perform numerical simulations of realistic laser pulses applied to the Kondo Lattice model. By tracking spin and charge degrees of freedom we identify a dynamical phase transition from a charge ordered into a metallic state, while preserving the spin order of the system. We propose to use higher harmonic generation to identify the dynamical phase transition in an ultrafast experiment.
\end{abstract}


\maketitle

\section{Introduction}

\begin{figure*}
\includegraphics[width=1.0\textwidth]{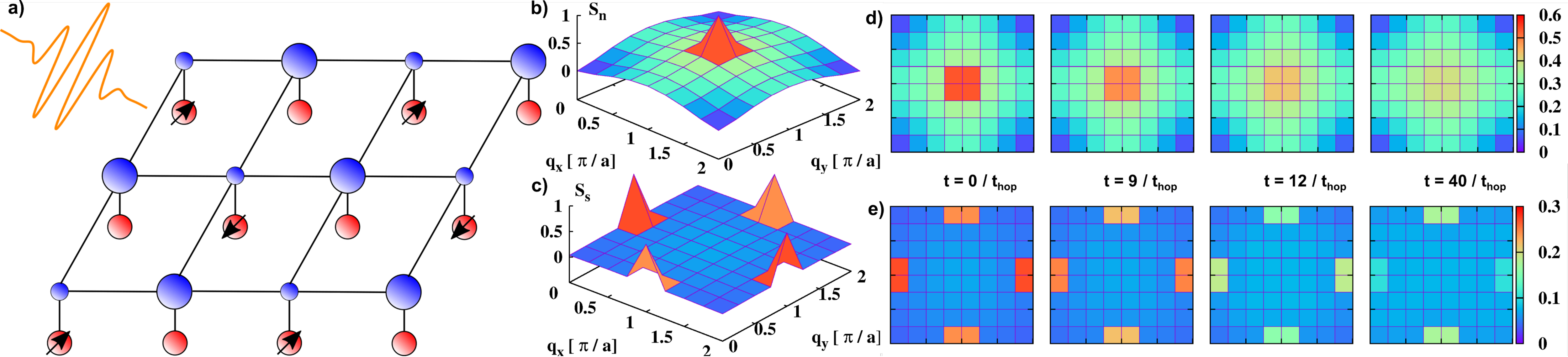}
\caption{\textbf{a)} Graphical representation of the ground state of the Kondo Lattice model for $J=3 t_\mathrm{hop}$ at quarter filling. Size of the blue circles is proportional to the charge density in the conduction band. Spin order is depicted for the $f$ electrons. \textbf{b)}  Momentum dependence of the static charge structure factor in the ground state. \textbf{c)} Momentum dependence of the static spin structure factor in the ground state. \textbf{d)} Charge structure factor time evolution for a pulse with peak amplitude $A_0 = 0.75$. \textbf{e)} Spin structure factor time evolution for a pulse with peak amplitude $A_0 = 0.75$. }
\label{Fig1}
\end{figure*}

Heavy fermion systems are one prototypical class of strongly correlated materials. 
The most important microscopic mechanism in these systems is the Kondo effect, which is determined by the strength of the local Kondo coupling. It can be tuned by conventional means, such as external or chemical pressure \cite{Wirth2016}. At the same time, the Kondo coupling also leads to an effective interaction, known as the Ruderman-Kittel-Kasuya-Yosida (RKKY) interaction, between the local moments mediated via the conduction band electrons.
The interplay between localized magnetic moments and conduction electrons are the driver behind unconventional superconductivity, competing phases and quantum criticality. As such, heavy fermion materials share commonalities with many other strongly correlated electron systems, such as cuprates, organic or iron-based superconductors. \\ 
Experimental advances in nonlinear optics and ultrafast spectroscopy allow for unprecedented access to nonequilibrium physics of these strongly correlated materials. By using this technique, many new and exciting phenomena have been discovered in recent years, such as light-induced superconductivity in cuprates \cite{Fausti189}, the generation of Higgs and order parameter oscillations \cite{PhysRevLett.111.057002, Werdehauseneaap8652, Schwarz2020}, and the dynamical coupling of ferroelectric and ferromagnetic order \cite{Sheu2014}. The search for new quantum phases, which can only be obtained by photoinduction, is an ongoing quest in modern solid state physics \cite{Kennes2017,Yang2018, Giorgianni2019}. It is guided by the idea of tailoring specific properties of materials on demand, for applications in electronic devices. \\
Although much of the progress focuses on nanostructures, complex oxides and oxide heterostructures (e.g., cuprates, manganites, and multiferroics), heavy fermion systems have been investigated in ultrafast experiments as well \cite{Wetli2018,PhysRevLett.104.227002,PhysRevB.97.165108}. However theoretical understanding of the transient non-equilibrium dynamics in these systems is still scarce. Many experiments can be explained by an effective temperature model \cite{PhysRevB.97.165108, PhysRevLett.99.197001,PhysRevB.85.144302}, which describes the local excitation by an instantaneous heating process. This approach is limited in describing the possible non-equilibrium phase transitions, because it is only sensitive to thermodynamic phases which are also present in equilibrium. \\
Here we use state-of-the-art numerical simulations to investigate the unitary time evolution of a driven Kondo system, by means of the time-dependent Variational Monte Carlo method \cite{Carleo2012,PhysRevA.89.031602,PhysRevB.92.245106,Idoe1700718,Carleo602}. Specifically we investigate the two dimensional Kondo Lattice model, which is used to model heavy fermion physics, at quarter filling. Studies by equilibrium Variational Monte Carlo \cite{PhysRevLett.110.246401} and Dynamical Mean Field Theory \cite{PhysRevB.92.075103} have confirmed the presence of a coexisting charge density and spin order in this system. A graphical representation of the ground state is given in Fig.\ \ref{Fig1}a). Note that coexisting charge and spin order has been reported in rare-earth intermetallic compounds \cite{PhysRevLett.85.158, PhysRevB.95.235156}. By shaking the ground state with a strong laser field, we investigate the stability of both order parameters depending on the pulse intensity. We show, that a dynamical phase transition can be induced by the pulse, leading to a purely spin ordered phase, with suppressed charge order. This phase is absent in the thermodynamic phase diagram, and can therefore not be captured by an effective temperature model. Analysis of the underlying many-body dynamics shows, that the transient electronic structure is strongly modified during the pulse, leading to a dynamical closure of the charge gap. Finally we suggest to use higher harmonic generation (HHG) spectroscopy as an experimental signature for the dynamical phase transition. 

\section{Model and Method}
The Kondo lattice model is defined by
\begin{align}
H = -t_\mathrm{hop}  \sum\limits_{\langle i , j \rangle, \sigma} \left( c_{i \sigma}^\dagger c_{i \sigma}^{\phantom\dagger} + \mathrm{H.c.} \right) + J \sum\limits_{i} \mathbf{S}_i^\mathrm{c} \cdot \mathbf{S}_i^\mathrm{f} .
\end{align}
The first term is the kinetic energy of the conduction electrons; the second term is the Kondo coupling of the conduction band to the local $SU(2)$ moments $\mathbf{S}_i^\mathrm{f}$ of the $f$-electrons. In the following we use $e=c=\hbar=a=1$, where $a$ is the lattice spacing. Unless otherwise noted, we use the hopping $t_\mathrm{hop}$ as our unit of energy and $1/t_\mathrm{hop}$ as our unit of time. \\
Focusing on the quarter filled case, the electron density is given by $n = 1/2$. In our simulations we choose a square lattice of size $8 \times 8$ with periodic (antiperiodic) boundary conditions in $x$ ($y$) direction. Note that larger systems can be simulated for pure ground state properties. The time evolution however demands much more computational effort and is therefore the restricting factor.\\
We use Variational Monte Carlo (VMC) and its time dependent version (tVMC) to compute ground state and time-evolved properties of the system, respectively.
The variational wave function is given by
\begin{align}
\left| \Psi \right\rangle = P_G P_J \left| \phi \right\rangle, \quad  \left| \phi \right\rangle = \left( \sum\limits_{\alpha, \beta =  \langle c, f \rangle } \sum\limits_{i,j}^{N_\mathrm{s}} f_{ij} \alpha_{i \uparrow}^\dagger \beta_{j \downarrow}^\dagger \right)^{N_e/2} \left| 0 \right\rangle .
\end{align}
Here $\left| \phi \right\rangle$ is the Pfaffian wave function for conduction band and $f$ electrons. The correlation factors $P_G$ and $P_J$ are of Gutzwiller and Jastrow type. For $f$ electrons the Gutzwiller factor is the only relevant correlation factor, as it permits only single occupancy of the lattice sites, effectively casting the $f$ degrees of freedom to a single spin-$1/2$. The variational parameters are subject to a $2 \times 2$ sublattice structure \cite{MISAWA2019447}. \\
The time-dependent EM field is included by the well established Peierls substitution \cite{Claassen2017,Konstantinovaeaap7427,PhysRevLett.112.176404}
\begin{align}
t_\mathrm{hop} \rightarrow t_\mathrm{hop} e^{i \mathbf{A}(\tau) (\mathbf{r}_i - \mathbf{r}_j)  },
\end{align}
where $\mathbf{A}(t)$ is the time-dependent vector potential. We choose a diagonal polarization $\mathbf{A}(t) = A(t) (1,1)^T$ and parameterize the pulse according to $A(t) = A_0 \exp( -(t - t_\mathrm{c})/(2 t_{\mathrm{d}}^2) ) \cos(\omega t)$. Here $A_0$ is the pulse amplitude, $t_\mathrm{c}$ the center, $t_{\mathrm{d}}$ the width and $\omega$ the frequency of the pulse. The pulse parameters that we use are, $A_0 = 0.5, 0.75, 1.0$, $\omega = 1.0 t_\mathrm{hop}$, $t_\mathrm{c} = 15.0 / t_\mathrm{hop}$ and $t_{\mathrm{d}} = 5.0 / t_\mathrm{hop}$.
Note that these parameters are well within the reach of experimental setups. Taking the lattice spacing from CeRhIn$_5$, in which a coexisting charge density wave with antiferromagnetic ordering was induced by a magnetic field \cite{Moll2015}, $a = 0.4656$nm \cite{Moshopoulou2002} and assuming a typical bandwidth of about $1$eV for the conduction electrons leads to a hopping parameter of $t_\mathrm{hop} = 0.125$ eV. Therefore the laser pulse frequency is in the range of $~15 - ~30$ THz with a the pulse width of $~52$ fs. The laser amplitude corresponds to a maximal electric Field $E_0 \propto 26.6$ MV/cm.

\section{Results}

\paragraph{Ground state properties at quarter filling}

To characterize the ground state properties, we compute the spin and charge structure factors
\begin{align}
S_{S} (\mathbf{q}) &= \frac{1}{N_\mathrm{S}} \sum\limits_{i,j} \left\langle \mathbf{S}_i \cdot \mathbf{S}_j \right\rangle  e^{i \mathbf{q} (\mathbf{r}_i - \mathbf{r}_j)} \\
S_{N} (\mathbf{q}) &=  \frac{1}{N_\mathrm{S}} \sum\limits_{i,j} \left\langle (n_i - \left\langle n_i \right\rangle) (n_j - \left\langle n_j \right\rangle) \right\rangle e^{i \mathbf{q} (\mathbf{r}_i - \mathbf{r}_j)} ,
\end{align}
for $J=3 \, t_\mathrm{hop}$ at the quarter filling. They are shown in Fig.\ \ref{Fig1}b) and c), respectively. We see strong peaks in the spin structure factor at $q=(0, \pi)$ and $q=(\pi,0)$ (and equivalently at  $q=(\pi,2 \pi)$ and  $q=(2 \pi, \pi)$)), indicative of antiferromagnetic order, which is not of the conventional N\'{e}el type. Note that the peak height is slightly different for $x$ and $y$ direction, due to the different boundary conditions. The charge structure factor shows a single strong peak at $q=(\pi,\pi)$, corresponding to a checkerboard type charge order. The unusual spin order leads to a magnetic unit cell, which is 4 times larger than the geometric unit cell. Note that the charge ordered phase is an insulator, with a charge gap $\Delta_c \approx 0.4 \, t_\mathrm{hop}$ at $J=3 \, t_\mathrm{hop}$ \cite{PhysRevLett.110.246401}. \\

\begin{figure}
\includegraphics[width=0.9\columnwidth]{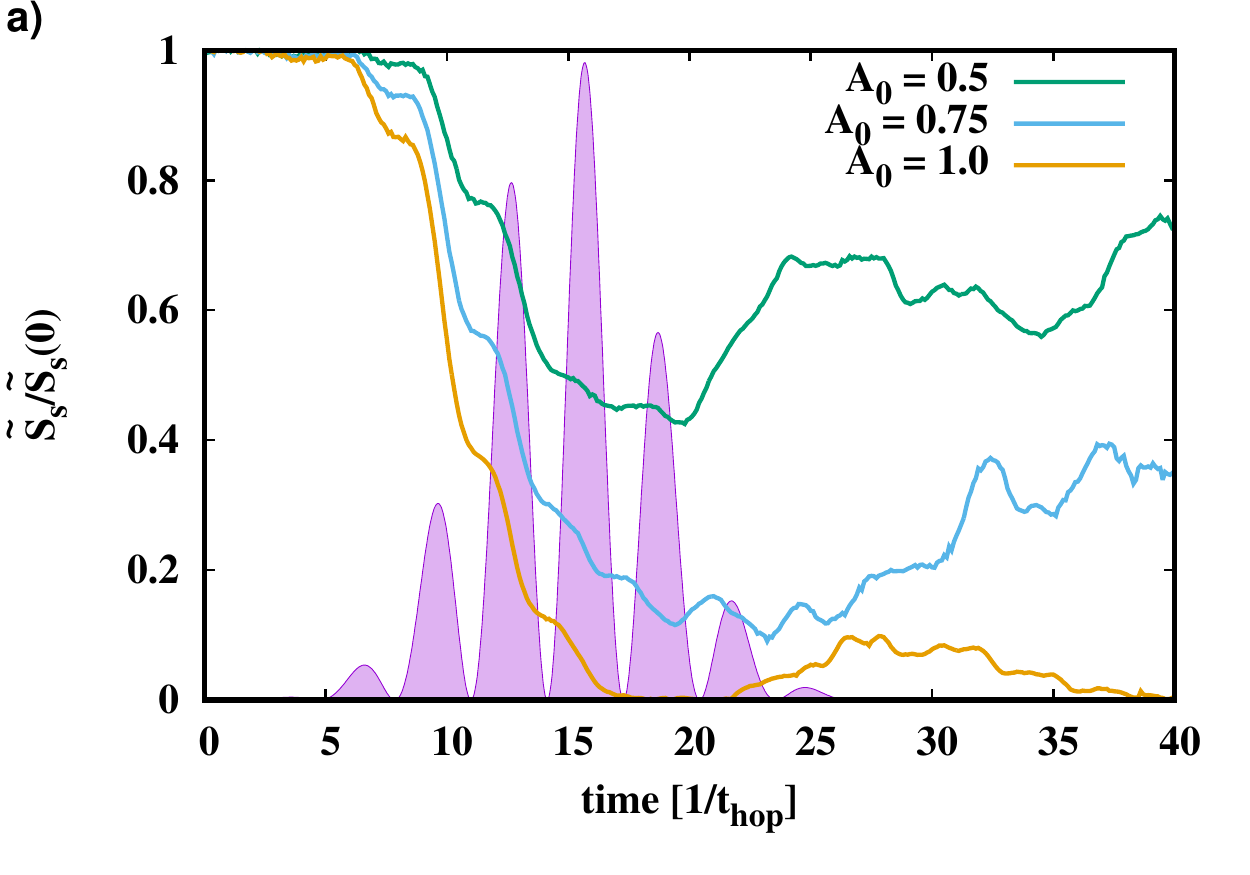}
\includegraphics[width=0.9\columnwidth]{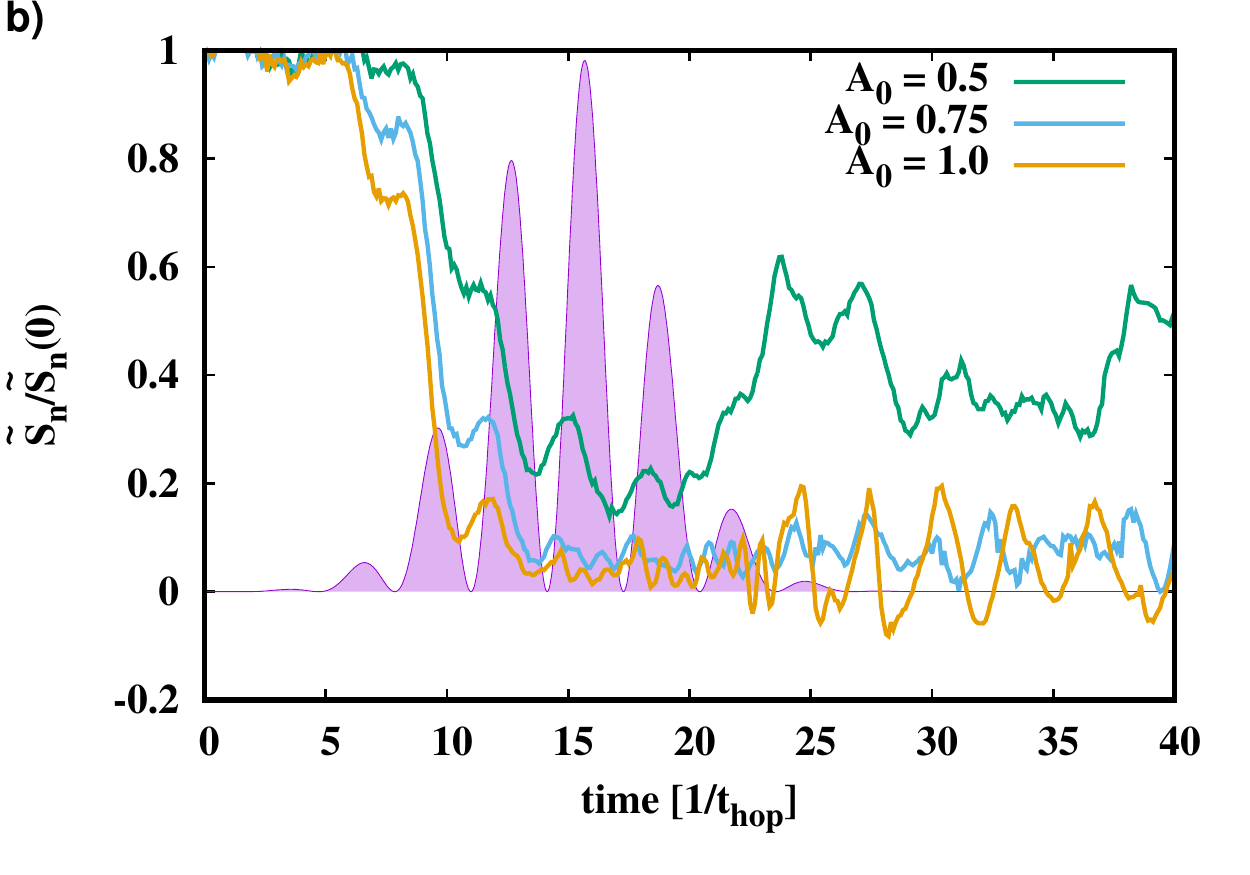}
\caption{\textbf{a)} Time evolution of the peaks in the spin static structure factor for different laser pulse amplitudes. The shaded background is proportional to the squared pulse profile. \textbf{b)} Time evolution of the peak in the charge structure factor for the same parameters as in \textbf{a)}. }
\label{Fig2}
\end{figure}

\paragraph{Time evolution}  

\begin{figure}
\includegraphics[width=0.9\columnwidth]{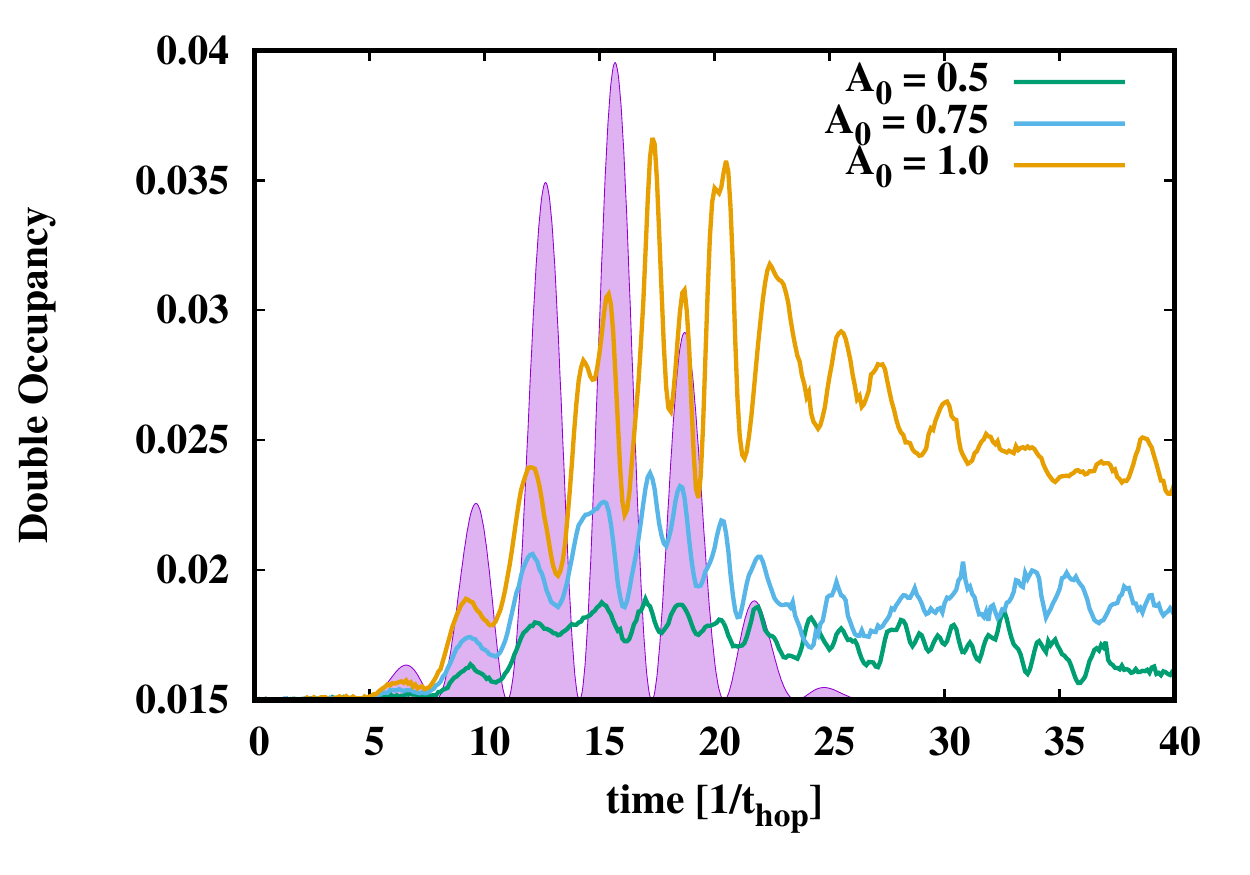}
\caption{Time evolution of the double occupancy of the conduction electrons for different laser pulse amplitudes. The shaded background is proportional to the squared pulse profile.}
\label{Fig3}
\end{figure}

\begin{figure}
\includegraphics[width=0.9\columnwidth]{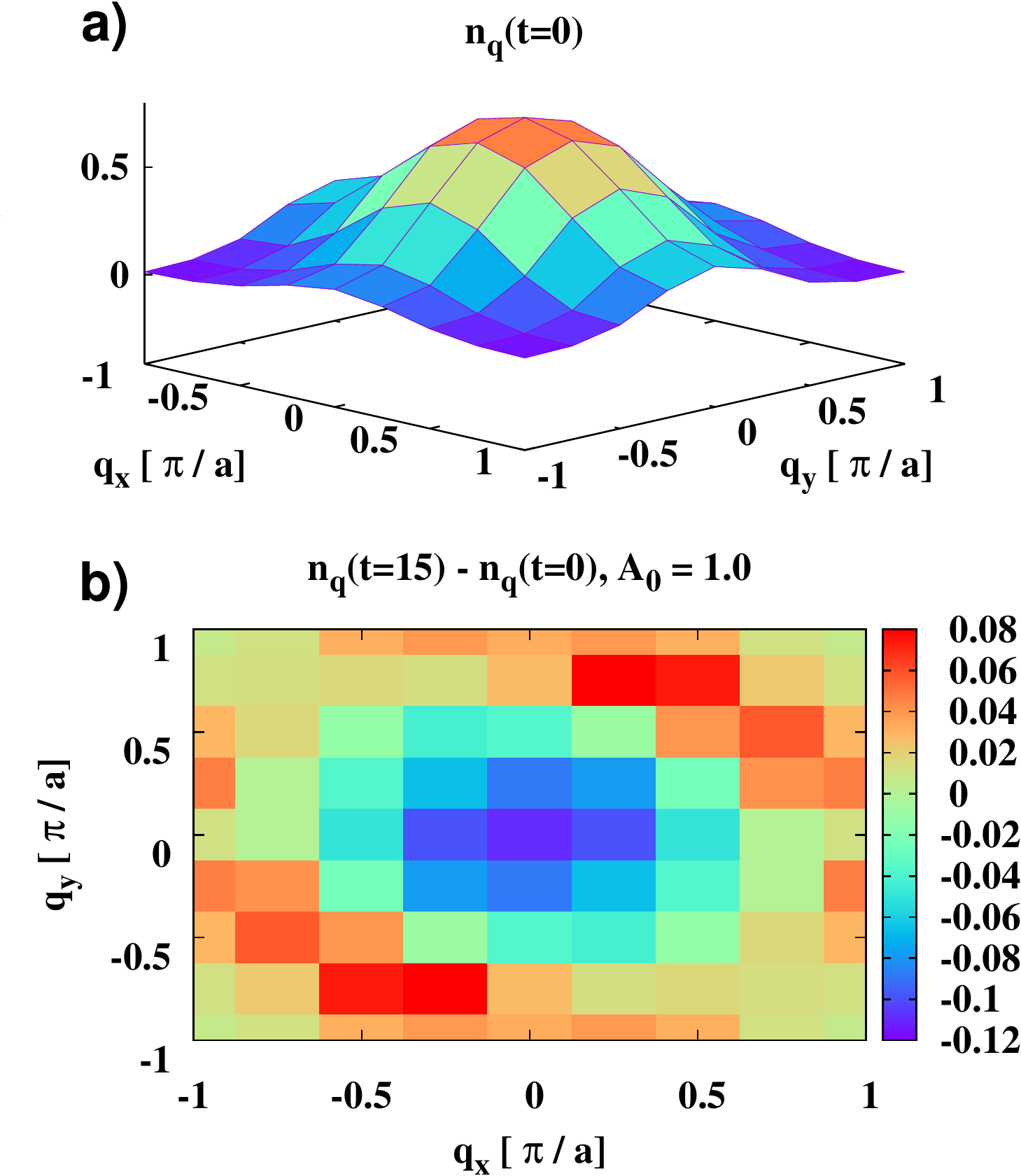}
\caption{\textbf{a)} Momentum distribution of the conduction electrons in the ground state of the Kondo Lattice model for $J=3\mathrm{t}$ and quarter filling. \textbf{b)} Difference between the Momentum distribution in the ground state and at $t=15$ for a laser pulse with $A_0=1.0$.}
\label{Fig4}
\end{figure}

The system is perturbed with a laser pulse and we track the dynamics in the spin and charge sectors. An overview of the time evolution of the spin and charge structure factors is shown in Fig.\ \ref{Fig1}d) and e), for a pulse amplitude of $A_0 = 0.75$. As the pulses progesses, the peaks in the charge and spin order are strongly suppressed by the pump pulse. The remaining Brillouin zone remains largely unaffected, indicating that the pulse primarly modifies the dominating charge and magnetic fluctuations. To make more quantitative statements the time dependence of the peaks is investigated.
We therefore define the relative height of the peaks in the charge and spin structure factor as,
\begin{align}
\tilde{S}_\mathrm{s}(t) &= \frac{S_\mathrm{s}((0,\pi),t) +
S_\mathrm{s}((\pi,0),t)}{2} - \underset{q \in \tilde{q}}{\mathrm{mean}} [S_\mathrm{s}(\mathbf{q},t)] ,\\
\tilde{S}_\mathrm{n}(t) &= S_\mathrm{n}((\pi,\pi),t) - \underset{q \in \tilde{q}}{\mathrm{mean}} [S_\mathrm{n}(\mathbf{q},t) ] ,
\end{align}
where $\tilde{q}$ are all momentum space points in direct neighborhood to the peak positions in the Brillouin zone. We then compare the effect of different pulse amplitudes. The results are given in Fig.\ \ref{Fig2}. \\
Our first observation is, that the pulse suppresses both peaks during the time evolution. Increasing the pulse amplitude in general leads to a stronger suppression. The time domain in which the peak height shows the strongest decrease coincides with the maxima in the pulse line shape (see shaded regions in Fig.\ \ref{Fig2}), reflecting the direct influence of the laser pulse onto the electronic and magnetic structure.  Depending on the response of the system, we can distinguish three different regimes: 1) For weak pulses, $A=0.5$, both charge and spin order are suppressed but stay finite after the pulse.  We see slow oscillations in the time domain, even after the pulse has ended. Note that low intensity high frequency oscillations are expected on top of the overall dynamics due to the finite system size. 2) In the intermediate regime at $A=0.75$ the charge order in Fig. \ref{Fig2}b) is strongly suppressed and on average close to zero, while the spin order is still larger than $30\%$ of the ground state value. 3) Further increasing the intensity to $A=1.0$ also suppresses the spin order, while the charge order shows stronger oscillations but with an average close to zero. \\
These results are to be contrasted with the thermal phase diagram \cite{PhysRevLett.110.246401}, in which thermal fluctuation first suppress spin order, while the critical temperature for the charge order is much higher. The direct coupling of the EM field to the charge is a possible explanation for this behavior: while the spin degrees of freedom of the $f$ electrons are only indirectly affect through the Kondo coupling, the conduction band electrons react directly to laser pulse. In contrast thermal fluctuations do not distinguish between charge and spin sector. \\
Note that the suppression of the peaks in the charge and spin sector correspond to a suppression of the order parameters in the thermodynamic limit. Thus our simulations directly hint at a dynamic phase transitions in time, controllable by the pulse amplitude. \\
To get a deeper microscopic understanding of the dynamics, we also calculated the time evolution of the double occupancy and the momentum distribution,
\begin{align}
n_\mathbf{q}(t) = \frac{1}{2 N_\mathrm{s}} \sum\limits_{i,j,\sigma} \left\langle c_{i \sigma}^\dagger c_{j \sigma}^{\phantom\dagger} \right\rangle (t) e^{i \mathbf{q} (\mathbf{r}_i - \mathbf{r}_j)}.
\end{align}
The double occupancy, shown in Fig. \ref{Fig3}, displays a strong oscillatory 
behavior during the pulse duration. While for the lowest pump amplitude the double occupancy almost recovers to its equilibrium value, higher pump amplitudes lead to a significant change and increase. Note that the value of the double occupancy for the metallic, non-interacting system is $1/16 = 0.0625$. Thus, judging only from the observed time evolution of the double occupancy, the system becomes more metallic for stronger pump pulses.\\
This observation is supported by the time evolution of the momentum distribution, shown in Fig.\ \ref{Fig4}. In equilibrium, Fig.\ \ref{Fig4}a), the momentum distribution has a smooth dependence on momentum, indicative of an insulating behavior. During the pulse evolution the momentum distribution stays smooth as a function of momentum, but weight is transfered away from the center of the Brillouin zone to the outer regions, see Fig.\ \ref{Fig4}b). This corresponds to the creation of free charge carriers and thus a breakdown of the charge insulator. \\
To further verify the dynamical melting of the charge insulator we investigate the small momentum dependence of the charge structure factor $S_{n} (\mathbf{q})$ in the supplemental material \cite{SuppMats}. We observe a change of the scaling from quadratic to linear, which is also observed for the equilibrium Mott transition in the Hubbard model \cite{PhysRevLett.94.026406, PhysRevLett.113.246405, 10.21468/SciPostPhys.7.2.021}. We conclude, that the dynamical phase transition is accompanied by a meltdown of the insulating state. \\
Note that our results do not rely on the specific form of the charge order. Thus we expect that the dynamical phase transition can also be observed at different filling factors, in which charge order can exist with a different wavelength \cite{PhysRevB.92.075103}.

\paragraph{Higher Harmonic Generation}

\begin{figure}
\includegraphics[width=0.9\columnwidth]{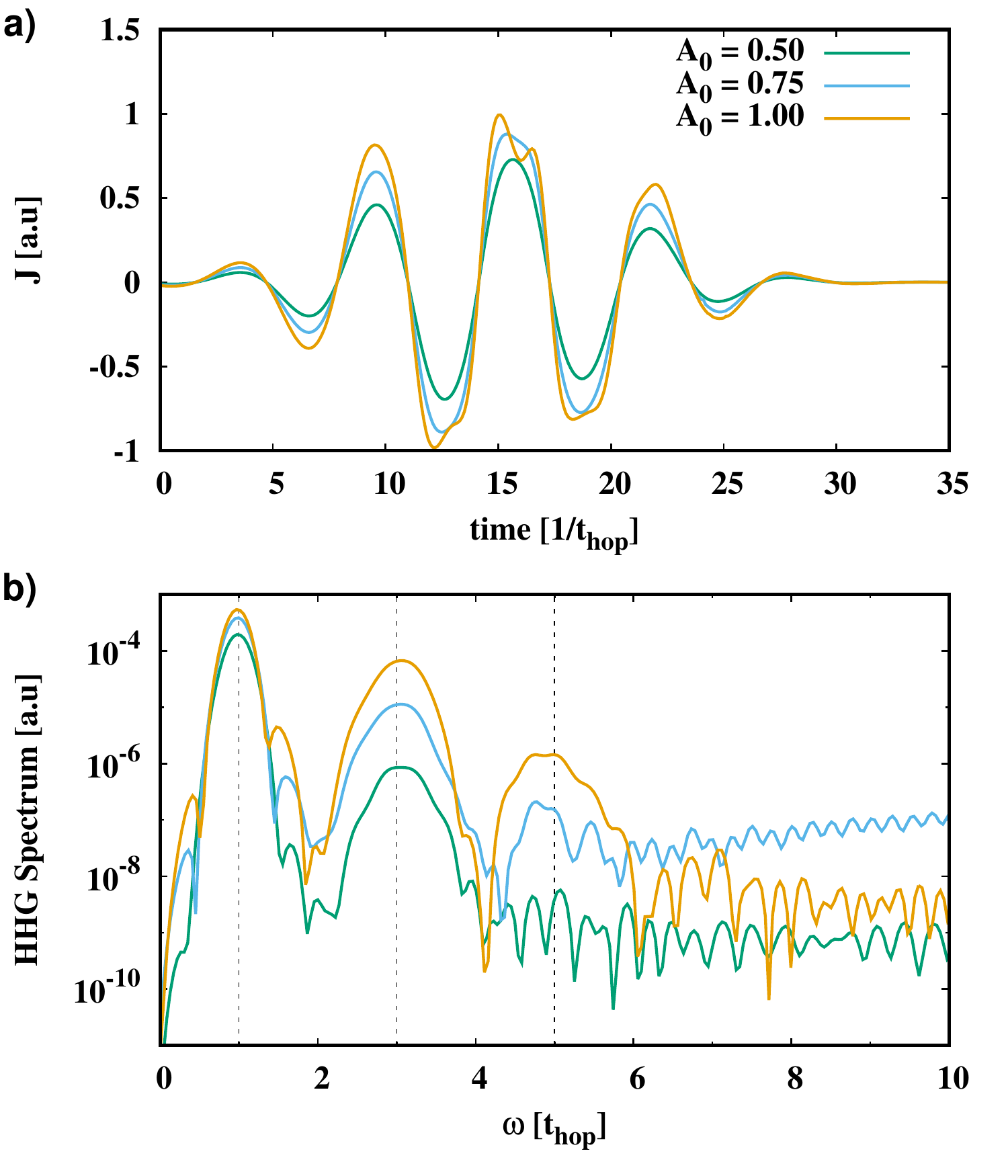}
\caption{\textbf{a)} Induced current for different laser pulse amplitudes. \textbf{b)} Higher Harmonic Generation spectrum for the same currents. Vertical lines show the first, third and fifth harmonic position.}
\label{Fig5}
\end{figure}

In a typical ultrafast experiment, a strong pump pulse
is used as an excitation, while a weaker probe pulse measures
the response of the system with a time delay. Here we show, that the single pump pulse is sufficient
to obtain insight into its effect on the system. We therefore compute the current, that is induced by the pump pulse. The result is shown in Fig.\ \ref{Fig5}a). For weak pump fluence, $A=0.5$ the response of the system is primarily linear in the EM-field, while we see clear non-linearities in the current for stronger fluence, e.g. $A=1.0$. We traced these non-lineareties back to the 
generation of higher harmonics, which can be observed in the 
current emission spectroscopy,
\begin{align}
FT\left[ \frac{\mathrm{d}J(t)}{\mathrm{d}t} \right](\omega),
\end{align}
where $FT$ denotes the Fourier transformation.
We observe that odd harmonics are generated by the pump pulse in the system.
It is expected, that photo excited charge carriers create a non-linear optical
response in solids \cite{Ghimire2011,Ghimire2019}. The lack of even harmonics is a consequence of the inversion symmetry of the system \cite{PhysRevA.56.748,Kemper_2013}.
While the fundamental harmonic is only weakly dependent on the pulse amplitude, higher harmonics
show a strong dependence on the pulse amplitude. In the third harmonic we see an increase in the amplitude
by two orders of magnitude by doubling the laser amplitude from $A=0.5$ to $A=1.0$. The effect in the fifth harmonic is even stronger: in the weak pulse regime the fifth harmonic cannot be distinguished from the background, while it shows a strong peak, once the charge and spin order are suppressed. Thus there seems to be a strong feedback effect on the higher harmonics depending on the internal dynamics of the charge and spin orders.
Besides measurements of optical conductivity, this could serve as an experimental indicator to identify the different dynamical regimes induced by the pump pulse.
Although we are only simulating a small system size, we expect that these results also hold in the thermodynamic limit, as the additional oscillations comming from finite size effects are small, as seen in the dynamics of the spin and charge structure factors and the double occupancy.\\ 
A possible explanation for the observed behavior is the dynamical phase transition from an insulating to a conducting state. In a simple model the creation of higher harmonics is a two step process, first charge carriers are excited from the ground state to a conduction band \cite{PhysRevA.85.043836}. Afterwards they are accelerated by the finite electric field, leading to an intraband current. Due to the strong field, electrons are Bragg scattered at the zone boundaries, leading to Bloch oscillations and radiation of photons with odd multiples of the laser driving frequency.
In the non-interacting case, the cutoff frequency for the appearance of higher harmonics depends linearly on the maximal amplitude of the laser field \cite{PhysRevA.85.043836} but in interacting systems or in the presence of disorder, this behavior is modified \cite{PhysRevLett.121.057405}. In our case, the strong modification of the electronic ground state due to the laser pulse, leads to a significant enhancement of the HHG spectrum.

\section{Conclusion}

In this paper we have investigated the dynamics of the spin and charge degrees of freedom in the 2D Kondo Lattice model driven by a strong laser pump field. The interplay between charge density and spin order can be strongly influenced by the light field, allowing us to dynamically manipulate the basic characteristics of the system. By tuning the laser intensity, we have shown, that charge fluctuations can be dynamically suppressed in favor of spin order, effectively inverting the temperature phase diagram of the model and allowing us to decouple the spin from the charge order. The dynamical phase transition is accompanied by a meltdown of the charge gap, as evidenced by microscopic observables such as double occupancy and momentum distribution. Finally we have demonstrated, that higher harmonic generation can be used as an experimental characterization tool for the dynamical phase transition. \\

Our work provides a mechanism for the ultrafast manipulation of coexisting charge and spin orders, offering a way to explore transient and hidden phases, which can not be attained by conventional methods, such as temperature or pressure. In the spirit of ``Quantum phases on demand'' \cite{Basov2017}, this provides a theoretical approach for the application of tailored ultra short laser pulses to strongly correlated materials.

Many questions are raised by our results, which are to be explored. While the system we investigated exhibits a coexisting charge and spin order, heavy fermion systems also display unconventional superconductvitiy and antiferromagnetism \cite{PhysRevLett.103.087001,PhysRevLett.101.177002, Mizukami2011} and coexistance of both, which could be tuned in a similar fashion. Additional the effect of electronic relaxation, quantum criticality and impact of coupling to lattice phonons are of interest.

\section{Acknowledgements} 
We thank Chen-Yen Lai and Stuart Trugman for fruitful discussions. This work was supported by the U.S. DOE NNSA under Contract No. 89233218CNA000001 via the LANL LDRD Program. It was supported in part by the Center for Integrated Nanotechnologies, a U.S. DOE Office of Basic Energy Sciences user facility, in partnership with the LANL Institutional Computing Program for computational resources. 

\bibliographystyle{prsty}

\end{document}